\documentclass{Interspeech}

\interspeechcameraready 
\usepackage[para]{manyfoot}
\usepackage{comment}
\usepackage{multirow}
\usepackage{xcolor}
\usepackage{svg}

\title{Vela: Scalable Embeddings with Voice Large Language Models for Multimodal Retrieval}

\author[affiliation={1}]{Ruofan}{Hu}
\author[affiliation={1}]{Yan}{Xia}
\author[affiliation={1}]{Minjie}{Hong}
\author[affiliation={2}]{Jieming}{Zhu}
\author[affiliation={2}]{Bo}{Chen}
\author[affiliation={1}]{Xiaoda}{Yang}
\author[affiliation={1}]{Minghui}{Fang}
\author[affiliation={1}]{Tao}{Jin}

\affiliation{}{Zhejiang University}{China}
\affiliation{}{Huawei Noah's Ark Lab}{China}

\email{jint\_zju@zju.edu.cn}


\makeatletter
\newcommand{\nonumberfootnote}[1]{%
  \def\@thefnmark{}
  \@footnotetext{#1}%
}
\makeatother

\begin{document}

\maketitle
\nonumberfootnote{Tao Jin is the corresponding author.}

\begin{abstract}
Multimodal large language models (MLLMs) have seen substantial progress in recent years. However, their ability to represent multimodal information in the acoustic domain remains underexplored. In this work, we introduce Vela, a novel framework designed to adapt MLLMs for the generation of universal multimodal embeddings. By leveraging MLLMs with specially crafted prompts and selected in-context learning examples, Vela effectively bridges the modality gap across various modalities. We then propose a single-modality training approach, where the model is trained exclusively on text pairs. Our experiments show that Vela outperforms traditional CLAP models in standard text-audio retrieval tasks. Furthermore, we introduce new benchmarks that expose CLAP models' limitations in handling long texts and complex retrieval tasks. In contrast, Vela, by harnessing the capabilities of MLLMs, demonstrates robust performance in these scenarios. Our code will soon be available.
\end{abstract}
\section{Introduction}
With the rapid advancement of MLLMs, there is an increasing need for embedding models that facilitate cross-modal retrieval. In the acoustic domain, while Contrastive Language Audio Pretraining (CLAP) models \cite{CLAP2023, laionclap2023, mei2024wavcaps, bai2024audiosetcaps, sun2024auto} have demonstrated impressive performance in text-audio retrieval by aligning acoustic and linguistic representations through contrastive learning. However, several limitations still remain in these works.

One major challenge lies in the excessive dependence on contrastive learning, which limits their ability to fully capture the nuances of each modality \cite{xiao2025grounding}. Moreover, the text encoders used are optimized for short captions, restricting their capacity to handle long and complex queries commonly encountered in practical scenarios \cite{fan2023improving, jang2024mate}. Together, these constraints result in models that are confined to coarse-grained alignment at the caption level, and demand a substantial amount of paired audio-text data to facilitate the contrastive learning process.

\begin{figure}[htbp]
    \centering
    \begin{minipage}{0.49\linewidth}
        \centering
        \includegraphics[width=\linewidth,clip, trim=35 75 28 75]{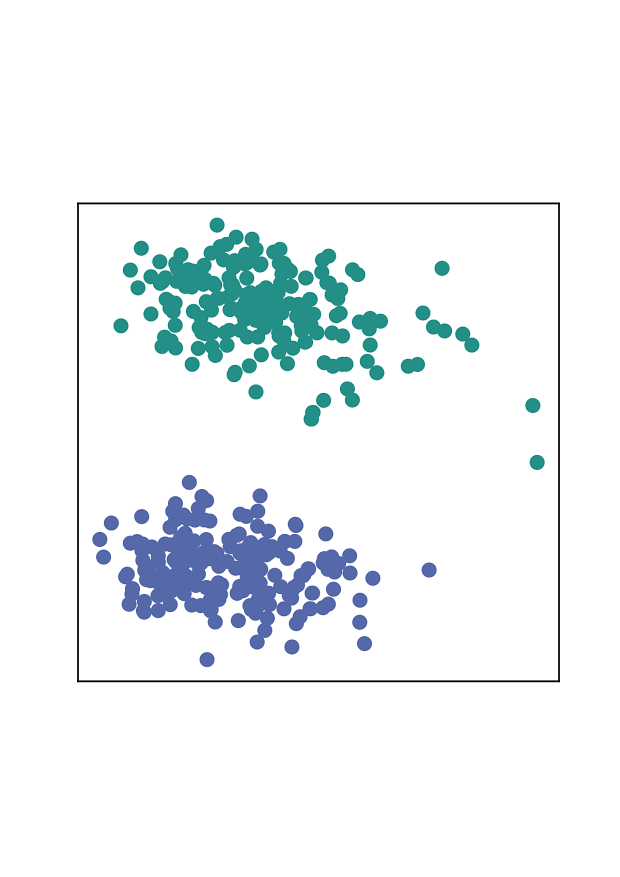}
        \caption*{(a) W/o. our method}
        \label{fig:first_image}
    \end{minipage}
    \hfill
    \begin{minipage}{0.49\linewidth}
        \centering
        \includegraphics[width=\linewidth, clip, trim=35 75 28 75]{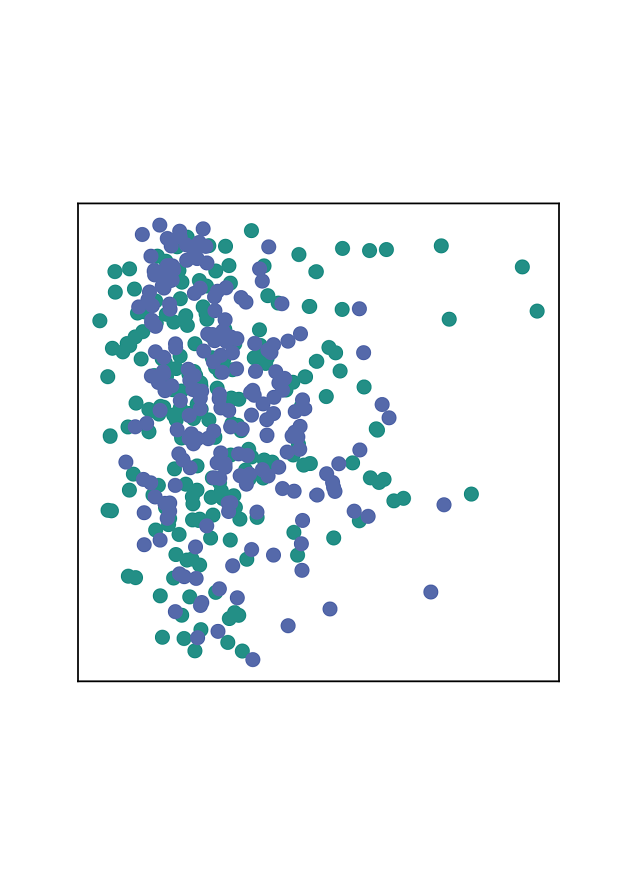}
        \caption*{(b) W/ our method}
        \label{fig:second_image}
    \end{minipage}
    \caption{Distribution of \textcolor[HTML]{238c84}{audio embeddings} and \textcolor[HTML]{5574aa}{text embeddings} from the last token in MLLMs. We compare the results with the ordinary prompt and with our representation method using the crafted prompt and in context learning design.}
    \label{fig:gapfig}
\end{figure}

Recent studies \cite{tong2024metamorph} highlight the importance of enhancing modality-specific understanding to improve generation performance. Based on these findings, we introduce Vela, a novel framework that employs MLLMs with enhanced modality comprehension for direct embedding generation. This approach offers several key advantages: First, unlike CLAP models, which depend on contrastive learning and are constrained by pairwise alignment, the multi-task paradigm allows MLLMs to deepen their understanding of multimodal semantics via cross-task knowledge transfer. Second, MLLMs benefit from advanced text encoders and external knowledge, enabling them to handle more complex and extended contextual sequences beyond the input constraints of CLAP. Third, MLLMs transfer their grounding and understanding of knowledge directly to the retrieval process, eliminating the need for additional paired audio-text data, which is challenging to collect.

Technically, we achieve this mechanism by guiding MLLMs to autonomously compress modalities, exploiting their autoregressive nature to derive representations from the token. However, the inherent modality gap between text and audio processing in MLLMs can adversely affect the alignment, as shown in Figure \ref{fig:gapfig}(a). To address this, we develop an efficient prompt engineering and in-context learning design that unifies cross-modal representations into a shared space \cite{fahim2024its, jiang2024e5} in Figure \ref{fig:gapfig}(b). Subsequently, we conduct training within this unified space under single-modality conditions. Our approach maintains frozen parameters across all MLLM components, while solely fine-tuning the language model through contrastive learning with positive-negative text pairs to strengthen its discriminative capabilities. Our contributions are:


\begin{figure*}[htbp]
    \centering
    \includegraphics[page=1, width=0.95\textwidth, clip, trim=48 65 11 70]{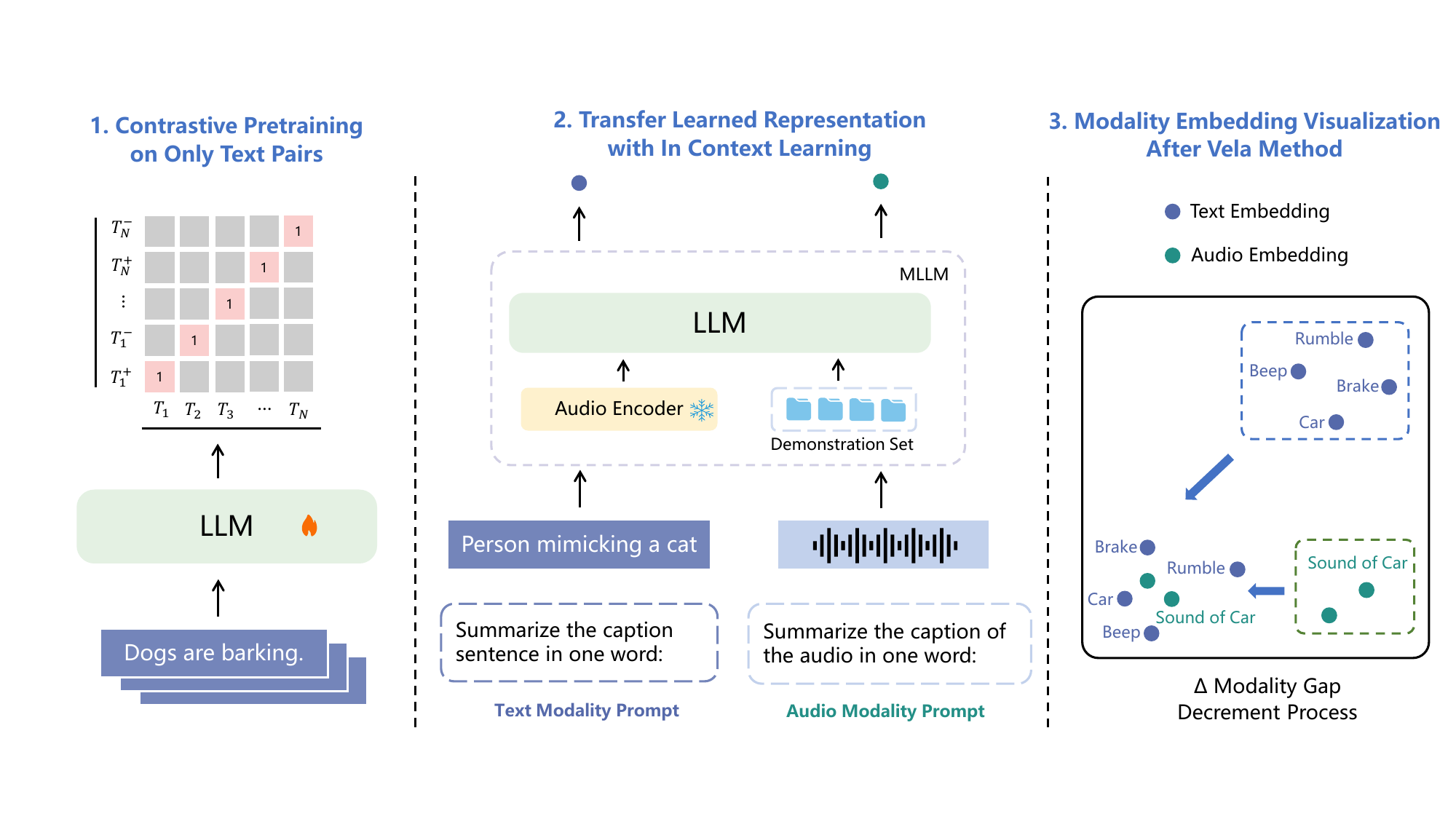}
    \caption{The whole framework of Vela. (1) In the \textbf{training} stage, by unifying multimodal representations into the same embedding space, Vela improves multimodal embeddings using \textbf{only contrastive learning on text pairs}. (2) In the \textbf{inference} stage, we freeze the modality encoder and projector in MLLM and replace the LLM with the pertaining one. With our representation method, embeddings are extracted. (3) Through the representation method and training of Vela, modality representations with similar semantics will aggregate together, which is further visualized using PCA.}
    \label{fig:interspeech}
\end{figure*}

\begin{itemize} 
\item We propose a method for generating universal multimodal embeddings using MLLMs, harnessing prompt-based design and in-context learning to activate retrieval task capabilities that are not natively supported by the original model.

\item Our framework outperforms CLAP-based methods that rely on text-audio pairs in retrieval tasks, even when trained on a limited number of text pairs in a single-modality setting.

\item We introduce novel acoustic benchmarks for retrieval tasks, highlighting the limitations of traditional CLAP methods and showcasing the advantages of our Vela over them.
\end{itemize}
\section{Method}
\subsection{Unifying Multimodal Embeddings}
Inspired by previous work \cite{jiang2024e5}, we adopt the ``\textbf{in one word}'' method with MLLMs to unify multimodal embeddings. The core idea behind is to guide MLLMs in compressing the information, and then leveraging the embeddings of the last token as the representation. Through extensive experimentation, we find that using prompts like \textit{\textless text\textgreater Summarize the caption sentence in one word:} and \textit{\textless audio\textgreater Summarize the caption of the audio in one word:}
 achieves the best results. This design is driven by two key considerations. First, it guides autoregressive MLLMs to generate a one-word summary, with its hidden state reflecting the input's probability distribution across the modalities. Second, explicitly incorporating terms like ``caption'' focuses the MLLM on the summarization task, thereby leveraging its multitasking capabilities to enhance the performance.

Additionally, previous studies have demonstrated that multimodal representations generated by the CLAP architecture exhibit a \textbf{modality gap} \cite{fahim2024its, radford2021learning, liang2022mind}, which negatively affects the alignment in retrieval tasks. While our prompts effectively generate semantically compressed embeddings, the modality gap remains unresolved. This occurs because, during semantic compression, MLLMs tend to prioritize words with higher syntactic salience in their conditional probability distributions, rather than those semantically critical for modality alignment. For instance, in a sentence like ``A person mimicking a cat’s meow'', the MLLM might emphasize ``mimicking``, whereas the focus should clearly be on ``meow''. This lexical salience bias ultimately perpetuates the modality gap despite effective information compression.

To address this, we draw inspiration from \textbf{in-context learning} \cite{tsukagoshi2021defsent}, which injects modality alignment priors into MLLMs. Specifically, our approach utilizes the \texttt{SoundBible\_flac} dataset \cite{mei2024wavcaps}, carefully selected for its acoustically distinctive samples containing singular sound events. To construct high-quality teaching exemplars, we employ a two-stage refinement process: (1) GPT-4 \cite{achiam2023gpt} is used for the initial one-word summary generation, followed by (2) Gemini \cite{team2023gemini}, which provides semantic fidelity scoring between the one-word summaries and the original captions, retaining the top 200 examples. In order to balance input length for autoregressive models with generation performance \cite{geng2024breaking}, we finalize the selection by keeping 65 text sentences and 35 audio clips as teaching exemplars.

\subsection{Single Modality Training with Contrastive Learning}
By unifying multimodal embeddings, we propose a novel single-modality training paradigm as illustrated in Figure \ref{fig:interspeech}. With the modality gap eliminated in the embeddings, this approach enables the transfer of single-modality representation capabilities to multimodal domains through contrastive learning on text pairs alone. To adapt to this modality-agnostic training framework, we streamline the architecture by freezing the modality encoder and projector, while training only the LLM component of the MLLM. We then utilize a simple NLI dataset \cite{gao2021simcse}, where each sentence pair $(x_i, x_i^+, x_i^-)$ has a positive sample $x_i^+$ and a negative sample $x_i^-$ for the input sentence $x_i$. We apply the same prompt and in-context learning examples for text modality to embed the sentence pairs into $(h_i, h_i^+, h_i^-)$. The training objective is as follows:


\[
\resizebox{\linewidth}{!}{$
\mathcal{L} = -\log \frac{\exp (\cos (h_i, h_i^+)/\tau)}{\sum_{j=1}^N \left( \exp (\cos (h_i, h_j^+)/\tau) + \exp (\cos (h_i, h_j^-)/\tau) \right)}
$}
\]

where $\tau$ is the temperature hyperparameter and $N$ is the batch size in contrastive learning. 

\subsection{Complex and Longer Benchmark}
The traditional CLAP framework is limited to coarse-grained, holistic representations and typically focuses on tasks involving short texts. In contrast, the Vela paradigm extends retrieval capabilities to long-text-to-audio tasks and complex retrieval tasks with designed instructions. To demonstrate this, we introduce two new benchmarks: Vela-long and Vela-conditional.

In \textbf{Vela-long}, we use an LLM \cite{achiam2023gpt} and Clotho \cite{drossos2020clotho} to expand the five captions into longer sentences while preserving their original meaning, increasing the average text length to \textbf{172.4} words—significantly more than the average of \textbf{11.32} words in the raw Clotho.

In \textbf{Vela-conditional}, we generate new samples by mixing diverse sounds \cite{mei2024wavcaps, fonseca2021fsd50k} and craft paired, narrative instructions that specify the target feature. For example, one new clip may include violin sounds, vehicle noises, and human speech, with the instruction guiding the model to ignore irrelevant elements and focus solely on the music. This methodology substitutes conventional comprehensive captions with succinct, directive instructions that emphasize partially features rather than overarching contextual elements. Each sound appears only once in the benchmark. We list all prompts in Table \ref{tab:prompt table}.

\begin{table}[ht]
\centering
\caption{All the prompts we use in the benchmark.}
\renewcommand{\arraystretch}{1.1} 
\label{tab:prompt table}
\begin{tabular}{p{1.3cm}|p{5.5cm}}
    \hline
    \textbf{Sources} & \textbf{Prompts} \\
    \hline
    \textbf{Clotho} & You will be given five short textual descriptions that correspond to a single clip. Your task is to merge these five descriptions into a single, coherent paragraph. The paragraph should be a reorganization and expansion of the original descriptions without adding any new information or altering the original meaning. Ensure the content is rich, but the meaning of each individual description must be preserved. The paragraph should flow naturally and read as a cohesive, well-structured piece of text. Make sure the final output is a smooth, connected paragraph that incorporates all five descriptions in a meaningful and coherent way. \\
    \hline
    \textbf{FSD50K} \textbf{Wavcaps} & You will be given one key label and three useless labels, you should design the prompt to instruct the system to search for the main label but ignore the useless labels.\\
\end{tabular}
\end{table}

\section{Experiment}

\subsection{Datasets}

For the training data, we use NLI \cite{gao2021simcse}, which contains approximately 273k sentence pairs. For the test data, we conduct text-audio retrieval on two open-source datasets, AudioCaps \cite{kim2019audiocaps} and Clotho \cite{drossos2020clotho}. To show the limitations of the CLAP structure in long-text and fine-grained retrieval tasks with instructions, we employ Vela-conditional, built on the FSD50 and Wavcaps \cite{mei2024wavcaps,fonseca2021fsd50k}, with approximately 400 examples, and Vela-long, based on Clotho, which includes around 1000 examples.

\subsection{Experimental Setup}

For the Vela backbone, we use Qwen2-Audio-7B-Instruct \cite{chu2024qwen2}, which is based on the Qwen2 \cite{bai2023qwen} series and utilizes a frozen Whisper-v3 \cite{radford2023robust} as the audio encoder. Due to hardware constraints, we trained the model using 8 NVIDIA 3090 GPUs, with a batch size of 168, over approximately 4 days. To optimize GPU memory usage, we adopted the QLoRA \cite{dettmers2024qlora} paradigm, applying LoRA with $r = 32$, $\alpha = 8$, and dropout = 0.05, and added LoRA modules to the linear layers of all 4-bit quantized models.

\section{Result}

\subsection{Text-Audio Retrieval}

Text-audio retrieval involves searching for an audio clip or a caption in a database based on a query from another modality. In our experiments, we selected several strong baselines based on the CLAP paradigm, including Microsoft-CLAP \cite{CLAP2023}, LAION-CLAP \cite{laionclap2023}, WavCaps \cite{mei2024wavcaps}, AudioSetCaps \cite{bai2024audiosetcaps} and Audio-ACD \cite{sun2024auto}. We select the \textbf{zero-shot version} when available.

We report Recall@K (R@K) for K=1, 5, 10 for both audio retrieval and text retrieval in the table. Compared to these strong baselines, Vela, as a universal multimodal embedding model, achieves competitive performance on both the AudioCaps and Clotho datasets. However, our model shows slightly lower performance than LAION-CLAP in terms of Recall@10 on Audiocaps.  This can be attributed to the fact that AudioCaps is a subset of the LAION-CLAP training data, leading to a similar data distribution. Additionally, since Vela was trained solely on the text modality, its \textbf{zero-shot} nature contributes to a slight performance drop. This is further supported by the results on WavCaps, where the zero-shot version, despite being trained on a larger dataset, shows lower performance during testing than other CLAP models.

Moreover, MLLMs such as Qwen2-Audio are typically designed for Next-Token prediction tasks in audio understanding. We demonstrate that, by using task-specific prompts not included in the training data, these models can still effectively transfer single-modality representation capabilities to multimodal embeddings. This indicates that the model can leverage its understanding to boost its generative capabilities, aligning with the current trend of unified understanding and generation paradigm in multimodal models \cite{tong2024metamorph}.

\begin{table*}[ht]
    \centering
    \caption{Comparison of audio and text retrieval performance on Audiocaps test set and Clotho evaluation datasets.}
    \renewcommand{\arraystretch}{1.1} 
    \label{tab:retrieval table 1}
    \begin{tabular}{ccccccccccccc}
        \toprule
        \multirow{2}{*}{\textbf{Method}} & \multicolumn{6}{c}{\textbf{audio retrieval}} & \multicolumn{6}{c}{\textbf{text retrieval}} \\
        \cmidrule(lr){2-7} \cmidrule(lr){8-13}
         & \multicolumn{3}{c}{Clotho} & \multicolumn{3}{c}{Audiocaps} & \multicolumn{3}{c}{Clotho} & \multicolumn{3}{c}{Audiocaps} \\
        \cmidrule(lr){2-4} \cmidrule(lr){5-7} \cmidrule(lr){8-10} \cmidrule(lr){11-13}
         & R@1 & R@5 & R@10 & R@1 & R@5 & R@10 & R@1 & R@5 & R@10 & R@1 & R@5 & R@10 \\
        \midrule
        \multicolumn{11}{l}{\textbf{Contrastive Learning on audio-text pairs}} \\
        Wavcaps & 15.8 & 39.1 & 50.2 & 35.5 & 68.1 & 78.5 & 17.1 & 39.2 & 51.2 & 37.1 & 70.2 & 81.7 \\
        Auto-ACD & 15.5 & 40.4 & 48.9 & 33.1 & 67.8 & 83.6 & 14.3 & 34.2 & 47.6 & 32.5 & 65.8 & 82.8 \\
        AudioSetcaps & 12.3 & 31.2 & 42.9 & 39.6 & 72.5 & 85.1 & 11.7 & 30.8 & 42.1 & 37.5 & 72.8 & 84.6 \\
        LAION CLAP & 15.2 & 39.1 & 50.9 & 37.7 & 80.8 & \textbf{91.2} & 14.4 & 37.5 & 50.1 & 38.5 & 73.8 & 86.4 \\
        Microsoft CLAP & 15.7 & 42.9 & 52.6 & 29.6 & 63.4 & 78.1 & 16.2 & 38.9 & 51.4 & 18.4 & 48.2 & 64.6 \\
        \midrule
        \multicolumn{11}{l}{\textbf{Contrastive Learning only on text pairs}} \\
        Vela & \textbf{18.1} & \textbf{43.6} & \textbf{53.1} & \textbf{41.8} & \textbf{82.1} & 89.3 & \textbf{17.9} & \textbf{40.2} & \textbf{50.6} & \textbf{40.2} & \textbf{75.3} & \textbf{87.6} \\
        \bottomrule
    \end{tabular}
\end{table*}

\subsection{Retrieval on Lengthy Caption and Conditional Instruction}

Most existing methods focus on short captions, which highlights a gap in cross-modal retrieval between audio and long texts (e.g., lengthy captions or documents) \cite{jang2024mate}. To explore this unexplored area, we used LLMs \cite{achiam2023gpt} to expand the captions in the Clotho evaluation set, significantly increasing their length while preserving semantic integrity. As shown in Table \ref{tab:recall table long}, we draw two main conclusions. First, as evidenced by the comparison with Table \ref{tab:retrieval table 1}, most existing models experience performance degradation when confronted with long-text-to-audio retrieval tasks with an equal number of test samples from clotho. Microsoft-CLAP, in particular, \textbf{fails} to support such long-text retrieval tasks. This underscores the need to improve the text encoders, even when using well-established models like HTSAT-BERT. Additionally, since our paragraphs are composed of five useful captions, they inherently provide more descriptive information about the audio content. As a result, Vela, leveraging the text encoder of LLMs, achieves better retrieval performance compared to using shorter texts in Table \ref{tab:retrieval table 1}. This further demonstrates the effectiveness of the MLLM framework in the retrieval domain.

\begin{table}[ht]
    \centering
    \caption{Recall@K Comparison Across Methods on text retrieval under Vela-Long}
    \setlength{\tabcolsep}{10pt} 
    \renewcommand{\arraystretch}{1.1} 
    \label{tab:recall table long}
    \begin{tabular}{lccc}
        \toprule
        \textbf{Method} & \textbf{K=1} & \textbf{K=5} & \textbf{K=10} \\
        \midrule
        Wavcaps & 10.1 & 26.1 & 40.0  \\
        Auto-ACD & 11.4 & 27.2 & 41.9 \\
        AudioSetcaps & 9.1 & 26.9 & 37.8 \\
        LAION CLAP & 13.8 & 35.7 & 49.1\\
        Microsoft CLAP & / & / & / \\
        \midrule
        \textbf{Vela} & \textbf{16.9} & \textbf{40.4} & \textbf{55.3} \\
        \bottomrule
    \end{tabular}
\end{table}

Additionally, we believe that future cross-modal retrieval will increasingly focus on conditional tasks with instructions instead of captions \cite{li2024matching}. We also conducted experiments using Vela-conditional. Instead of relying on captions, we replaced them with instructions that guide the model to perform retrieval under more fine-grained conditions, such as: "Ignore irrelevant information as human talk and find samples with violin sounds." In contrast, traditional tasks would use a full caption sentence, such as: "In the midst of chaotic traffic noise, a man is speaking loudly while there is a soft violin playing." As shown in the Table \ref{tab:recall table 4}, nearly all models exhibit a weaker understanding of instruction-based inputs, leading to a performance drop even with fewer samples. In contrast, Vela achieves higher performance due to its stronger understanding and reasoning capability enabled by MLLMs, contributing to the superior performance in our benchmark.

\begin{table}[ht]
    \centering
    \caption{Recall@K Comparison Across Methods on text retrieval under Vela-Conditional}
    \setlength{\tabcolsep}{10pt} 
    \renewcommand{\arraystretch}{1.1} 
    \label{tab:recall table conditional}
    \begin{tabular}{lccc}
        \toprule
        \textbf{Method} & \textbf{K=1} & \textbf{K=5} & \textbf{K=10} \\
        \midrule
        Wavcaps & 7.1 & 19.4 & 29.7 \\
        Auto-ACD & 7.2 & 18.5 & 24.3 \\
        AudioSetcaps & 6.7 & 17.9 & 22.6 \\
        LAION CLAP & 7.7 & 19.4 & 24.6 \\
        Microsoft CLAP & 10.2 & 21.0 & 28.7 \\
        \midrule
        \textbf{Vela} & \textbf{13.4} & \textbf{24.5} & \textbf{33.2} \\
        \bottomrule
    \end{tabular}
\end{table}

\subsection{Ablation Study}

Finally, to validate the effectiveness of our paradigm design, we conducted an ablation study, as shown in Table \ref{tab:recall table 4}. We started with ``in one word'' prompt design and progressively incorporated in-context learning and contrastive learning separately into Vela. By utilizing these approaches, we demonstrated the gradual reduction of the modality gap \cite{liang2022mind}, proving that MLLMs can map multimodal content into a unified space and support our premise of single-modality training. Additionally, our findings indicate that in-context learning plays a more significant role than contrastive learning. This is because MLLMs alone do not have the capacity to capture the semantic characteristics across different modalities that are crucial for retrieval tasks. Only by designing effective samples can they achieve unified mapping for paired modalities.

\begin{table}[ht]
    \centering
    \caption{Recall@K Comparison on Ablation Study. The Gap score reflects the difference between different modalities.}
    \setlength{\tabcolsep}{7pt} 
    \renewcommand{\arraystretch}{1.1} 
    \label{tab:recall table 4}
    \resizebox{\linewidth}{!}{
      \begin{tabular}{lcccc}
        \toprule
        \textbf{Configurations} & \textbf{K=1}  & \textbf{K=10} & \textbf{Gap} $\downarrow$ \\
        \midrule
        Prompt without ``one word'' & 4.69  & 25.30 & 67.2 \\
        Prompt with ``one word'' & 12.8 & 41.7 & 44.1 \\
        ``One word'' prompt with samples & 13.6 & 47.1 & 28.9\\
        ``One word'' prompt with training & 7.04 & 34.3  & 40.6 \\
        \midrule
        \textbf{With all proposals} & \textbf{18.1} & \textbf{53.1} & \textbf{21.9} \\
        \bottomrule
    \end{tabular}
    }
\end{table}

\section{Conclusion}
In this work, we propose Vela, an MLLM-based universal multimodal framework capable of representing both acoustic and linguistic inputs. Technically, it employs a prompt-based representation design and an in-context learning approach to unify multimodal representations into a shared embedding space without fine-tuning. Through single-modality training on text pairs, Vela demonstrates strong performance on audio-language retrieval tasks. More importantly, by introducing our new benchmark, we highlight the limitations of existing CLAP-based models in long-text tasks and conditional retrieval tasks with instructions. We also conducted an ablation study to validate the effectiveness of our approach. We hope that our work can provide valuable insights for the development of unified multimodal representations in the future.
\section{Acknowledgements}
This work was supported by the ``Pioneer'' and ``Leading Goose'' R\&D Program of Zhejiang under (Grant No. 2025C02110), Public Welfare Research Program of Ningbo under (Grant No. 2024S062), and Yongjiang Talent Project of Ningbo under (Grant No. 2024A-161-G).



\bibliographystyle{IEEEtran}
\bibliography{mybib}

\end{document}